\documentclass[preprint,showpacs,aps]{revtex4}
\usepackage{graphicx}
\usepackage{dcolumn}
\usepackage{bm}
\usepackage{color}
\def\slash#1{/\hspace{-0.22cm}{#1}}
\def\tr{\mathop{\mathrm{Tr}}\nolimits}

\begin{document}

\title{Coleman-Weinberg mechanism for spontaneous chiral symmetry breaking in the massless chiral sigma model}

\author{Setsuo Tamenaga}
\email{stame@rcnp.osaka-u.ac.jp}
\author{Hiroshi Toki}
\email{toki@rcnp.osaka-u.ac.jp}
\author{Akihiro Haga}
\email{haga@rcnp.osaka-u.ac.jp}
\author{Yoko Ogawa}
\email{ogaway@rcnp.osaka-u.ac.jp}
\affiliation{Research Center for Nuclear Physics (RCNP), \\
Osaka University, Ibaraki, Osaka 567-0047, Japan
}

\date{\today}% It is always \today, today,

\begin{abstract}
We study the effect of one-loop corrections from
nucleon together with those from boson in the massless chiral sigma model, 
where we perform the Coleman-Weinberg renormalization procedure.
This renormalization procedure has a mechanism of 
spontaneous symmetry breaking due to radiative corrections
in $\phi^4$ theory.
We apply it to the system of nucleon and bosons 
with chiral symmetry where the negative-mass
term of bosons does not exist.
Spontaneous chiral symmetry breaking is derived from 
the contribution of nucleon and boson loops
which generates the masses of nucleon, scalar meson,
and vector meson dynamically.
We find that the renormalization scale plays
an important role for the breaking of the symmetry between fermion (nucleon) and
boson, and eventually for the chiral symmetry at the same time.
In addition, we find that the naturalness restores by means of the introduction of the vacuum fluctuation from both nucleon and boson. 
Finally we obtain a stable effective potential 
with the effect of Dirac sea in the chiral model
for the first time.
\end{abstract}

\maketitle

\section{\label{sec:level1}Introduction}
The relativistic mean field (RMF) approximation is very popular
now for the description of nuclei and nuclear matter.  The success
of the RMF model originates from the presence of large scalar and
vector potentials with opposite sign \cite{walecka}.  The scalar
potential is attractive and proportional to the scalar density,
while the vector potential is repulsive and proportional to the
vector density.  This fact makes the net potential attractive
for small densities and repulsive for large densities, the feature
of which provides the saturation property of nuclear matter. The
large scalar and vector potentials with opposite sign provide a
large spin-orbit potential, which is necessary to provide the
jj-closed shell magic numbers \cite{meyerjensen}. Further
introduction of non-linear terms of the sigma meson field with a
few additional parameters is able to make the incompressibility
appropriate and describe nuclei quantitatively in the entire mass
region \cite{rheinhard,sugahara,lalazisis}.

In principle, however, we have to include the contribution of
negative-energy states to the total energy and the densities due
to the modification of the negative-energy states caused by
formation of the nucleus.  This is because the nucleons with
positive energy polarize the negative-energy states and the whole
object including the negative-energy nucleons should be the
nucleus of our concern. In the development of the RMF model, we
have assumed that the parameters in the RMF Lagrangian mocks up
the contribution from negative-energy states.  In this sense,
the present RMF model is simply not more than a phenomenological
model. Definitely, there are several works to try to include the
contribution of negative-energy states \cite{walecka,greiner,nagata}.

A theoretical study of the RMF model of the nucleus 
was performed by using the sigma
model of Gell-mann and Levy \cite{gellmann}.  The use of the
original Lagrangian after making the Weinberg transformation in
addition to the introduction of the omega meson term was not
successful to generate a good saturation property of nuclear
matter in the mean field approximation \cite{boguta1}.  Hence, the
chiral model was modified by introducing the dynamical mass
generation term for the omega meson as the case of the nucleon.  This
modified sigma model, named as chiral sigma model, is able to
provide a good saturation property for nuclear matter
\cite{boguta1,ogawa1}. However, the incompressibility comes out to be too
large (K=650[MeV]), which causes a bad feature in the single
particle spectra of finite nuclei \cite{ogawa1}.

This observation sets a stage to study the modification of
negative-energy states, since the linear sigma model Lagrangian is
used for the hadron properties and for pion dynamics.  Hence, we
study the effect of the Dirac sea in the chiral sigma model for
nuclear structure. The chiral sigma model is renormalizable, and
it is legitimate to include the Dirac sea in this model. In the
non-chiral model (Walecka model) it is known that the contribution
of the Dirac sea is to make the scalar density decrease by about
10\% to 20\% at the saturation density and reduces the
incompressibility \cite{nagata}. We shall try now to include the
Dirac sea in the chiral sigma model. Since the chiral sigma model has the chiral symmetry,
the counter terms need to respect this symmetry \cite{matsui}. We
obtain the counterterms with the chiral symmetry,
but the counterterms remain arbitrary
and the total effective potential comes out
to be unstable \cite{boguta2}.
We should reconstruct one-loop corrections with chiral symmetry
using a new renormalization procedure
and the mechanism of the spontaneous symmetry breaking.  This is the purpose of this paper.

In Sect.~\ref{sec:level2}, we review the radiative corrections
as the origin of spontaneous symmetry breaking in the Coleman-Weinberg scheme \cite{coleman}.
Using the Coleman-Weinberg renormalization scheme, we calculate the one-loop corrections with chiral symmetry and construct the massless 
chiral sigma model with Dirac sea in Sect.~\ref{sec:level3}.
In Sect.~\ref{sec:level4} and \ref{sec:level5}, 
we show some properties of the massless chiral sigma model.
We summarize the present study in Sect.~\ref{sec:level6}.

\section{\label{sec:level2}Radiative corrections and 
spontaneous symmetry breaking}

\begin{figure}
\includegraphics[scale=0.6]{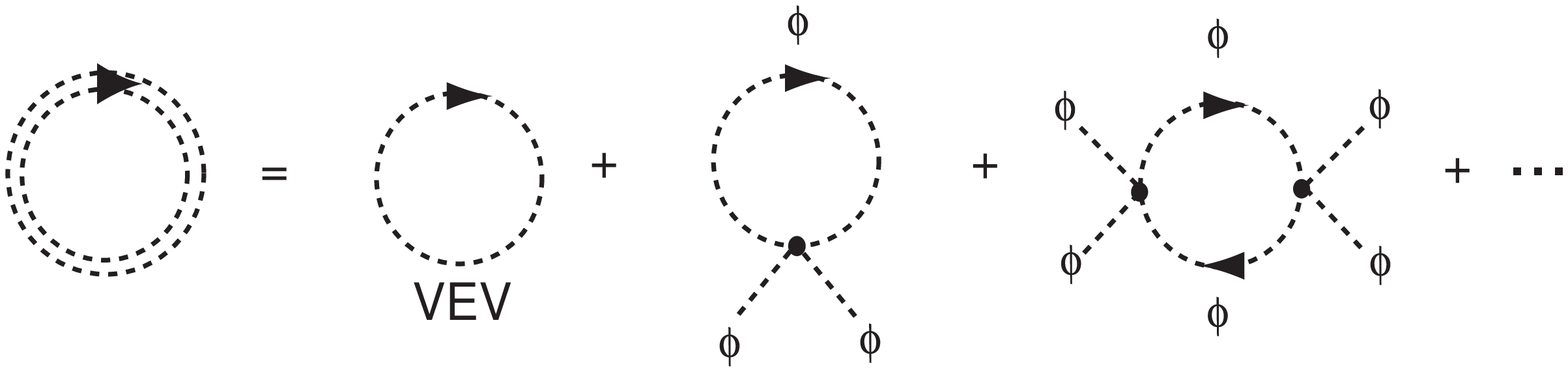}
\caption{\label{fig:fig1}All-order boson loop corrections
at the one-loop level.
First term of the right hand side is the vacuum expectation value (VEV)
corresponding to the contribution from the valence bosons.  The second and the third terms are the higher order corrections of VEV.}
\end{figure}

The chiral model usually has a mechanism of spontaneous
chiral symmetry breaking by construction. This mechanism comes from the negative-mass term of (Higgs) bosons
to make a minimum point in the wine bottle potential.
First, we review the radiative corrections of this system in the $\phi^4$ theory. The Lagrangian is
\begin{eqnarray}
\mathcal{L}=\frac{1}{2}(\partial_{\mu}\phi)^2-\frac{\mu^2}{2}\phi^2-\frac{\lambda}{4!}\phi^4-\delta\mathcal{L}_{CTC}.\label{eq:phi4}
\end{eqnarray}
This Lagrangian consists of only the field $\phi$ for scalar meson, and $\mu^2$, $\lambda$ are the
mass and the coupling constant. The last term $\delta\mathcal{L}_{CTC}$ is the counterterm Lagrangian,
which is necessary to renormalize the boson loop in Fig.~\ref{fig:fig1}. 
The effective action of Eq.~(\ref{eq:phi4}) at the one-loop
level can be written as
\begin{eqnarray}
\Gamma^B&=&\int d^4x\left[\frac{i}{2}\tr\ln\left(p^2-\mu^2-\frac{\lambda}{2}\phi^2\right)-VEV+\mathcal{L}\right]\nonumber\\
{}&=&\int d^4x\left[-V_B^R(\phi)+\frac{1}{2}\left(1+Z_{\phi}(\phi)\right)(\partial_{\mu}\phi)^2+Y_{\phi}(\phi)(\partial_{\mu}\phi)^4+\cdots\right]\label{eq:bosoncal}.
\end{eqnarray}
We take ordinary renormalization conditions for mass, coupling constant, and
derivative term to Eq.~(\ref{eq:bosoncal}) with counterterms,
\begin{eqnarray}
\left.\frac{\partial^2 V_B^R}{\partial\phi^2}\right|_{\phi=0}&=&\mu^2,\label{eq:masscon1}\\
{}\left.\frac{\partial^4 V_B^R}{\partial\phi^4}\right|_{\phi=0}&=&\lambda,\label{eq:couplecon1}\\
{}Z_{\phi}|_{\phi=0}&=&0.\label{eq:wavecon1}
\end{eqnarray}
Here, we keep only the lowest order derivative term, since the 
derivative expansion is known to converge rapidly \cite{greiner,fraser,haga}. 
Therefore we obtain the renormalized Lagrangian as,
\begin{eqnarray}
\mathcal{L}&=&\frac{1}{2}\left(1+Z_{\phi}\right)(\partial_{\mu}\phi)^2-V_B^R,
\end{eqnarray}
where
\begin{eqnarray}
V_B^R&=&\frac{\mu^2}{2}\phi^2+\frac{\lambda}{4!}\phi^4\nonumber\\
{}& &+\frac{\mu^4}{64\pi^2}\left[\left(1+\frac{\lambda\phi^2}{2\mu^2}\right)^2\ln\left(1+\frac{\lambda\phi^2}{2\mu^2}\right)-\frac{\lambda\phi^2}{2\mu^2}-\frac{3}{2}\left(\frac{\lambda\phi^2}{2\mu^2}\right)^2\right],\\
{}Z_{\phi}&=&\frac{\lambda^2\phi^2}{192\pi^2\left(\mu^2+\frac{1}{2}\lambda\phi^2\right)}.
\end{eqnarray}
This potential has a mechanism to break the spontaneous symmetry owing to the negative-mass term ($\mu^2< 0$).  This is the standard spontaneous symmetry breaking scheme. 

We introduce now the Coleman-Weinberg renormalization scheme \cite{coleman}.  In the Coleman-Weinberg scheme, we set the mass term vanish and let the radiative corrections have the same role as the negative-mass term. We change the renormalization conditions from (\ref{eq:masscon1}) $\sim$ (\ref{eq:wavecon1}) to the new ones as,
\begin{eqnarray}
\left.\frac{\partial^2 V_B^R}{\partial\phi^2}\right|_{\phi=0}&=&0,\\
{}\left.\frac{\partial^4 V_B^R}{\partial\phi^4}\right|_{\phi=m}&=&\lambda,\label{eq:lm}\\
{}Z_{\phi}|_{\phi=m}&=&0\label{eq:zm},
\end{eqnarray}
where we introduce the renormalization scale $m$ 
to Eqs. (\ref{eq:lm}) and (\ref{eq:zm}) 
in order to avoid the logarithmic singularity at the origin of the effective potential. Finally the renormalized effective potential becomes
\begin{eqnarray}
V_B^R&=&\frac{\lambda}{4!}\phi^4+\frac{\lambda^2\phi^4}{256\pi^2}\left[\ln\left(\frac{\phi^2}{m^2}\right)-\frac{25}{6}\right],\label{eq:rad}
%\\
%{}Z_{\phi}&=&\frac{\lambda}{96\pi^2}.
\end{eqnarray}
with the constant correction $\frac{\lambda}{96\pi^2}$ in $Z_{\phi}$ before the renormalization, which is removed by the counterterm.
Since the second term of Eq.~(\ref{eq:rad}) is negative around the origin, it has an effect to make a new minimum at some point away from the origin.  This mechanism plays the role of spontaneous symmetry breaking in the Coleman-Weinberg scheme.  This scheme is not used, however, in the actual case, since the radiative corrections are much bigger than the tree level contributions.

\section{\label{sec:level3}One-loop corrections with chiral symmetry}

It is well known that chirally symmetric renormalization in the conventional way gives unstable effective potential \cite{boguta2} and too large non-linear interactions \cite{furn}. These features come from the form of the counterterms and the renormalization conditions.
In order to respect the chiral symmetry of the model, the number of the counterterms is restricted to two terms despite of four divergent diagrams of the nucleon loop.
After spontaneous chiral symmetry breaking, we take only two
renormalization conditions for a local minimum and the mass
in the ordinary way \cite{matsui,lee}. In this way the coupling constant $\lambda$ could not be renormalized properly.

Boson-loop corrections were, then, introduced for the cancellation of the unstable effective potential coming from the nucleon loop \cite{jackson,lee}.  We have to take a large sigma meson mass as about 1125 MeV in order to achieve the cancellation of the nucleon and the boson loop corrections.  In this case, we get too small scalar potential and eventually too small spin-orbit splitting for the case of the nucleus.  In addition, the problem of instability is not solved perfectly.  Hence, until now, we do not have a satisfactory procedure to treat the vacuum polarization for the Lagrangian with the chiral symmetry.

We consider now the Coleman-Weinberg renormalization scheme
to both the nucleon and boson loops with chiral symmetry. As shown in the previous section, radiative corrections from boson are taken into account {\it before symmetry breaking} and give rise to the spontaneous symmetry breaking.  We consider then one-loop corrections for nucleon
{\it before the spontaneous chiral symmetry breaking} in the same way as the boson loop in the $\phi^4$ theory.

We begin with the chiral sigma model \cite{boguta1,ogawa1}:
\begin{eqnarray}
\mathcal{L}&=&\bar{\psi}\left[i\gamma_{\mu}\partial^{\mu}-
g_{\sigma}(\phi+i\gamma_5\bm\tau\cdot\bm\pi)-
g_{\omega}\gamma_{\mu}\omega^{\mu}\right]\psi\nonumber\\
{}&+&\frac{1}{2}\left(\partial_{\mu}\phi\partial^{\mu}\phi+
\partial_{\mu}\bm\pi\cdot\partial^{\mu}\bm\pi\right)-
\frac{\mu^2}{2}\left(\phi^2+\bm\pi^2\right)-
\frac{\lambda}{4}\left(\phi^2+\bm\pi^2\right)^2\nonumber\\
{}&-&\frac{1}{4}\Omega_{\mu\nu}\Omega^{\mu\nu}+
\frac{1}2{}\widetilde{g_{\omega}}^2\left(\phi^2+\bm\pi^2\right)
\omega_{\mu}\omega^{\mu}-\delta\mathcal{L}_{CTC},
\end{eqnarray}
where $\psi$, $\phi$, and $\bm\pi$ are nucleon field, sigma meson field, and pi meson field, respectively. This Lagrangian has the chiral symmetry.  The sigma meson field appears together with the pion field in the chiral symmetric way.  There does not exist the nucleon mass term, which is produced by chiral symmetry breaking in this chiral sigma model.  We introduce omega meson field, $\omega_{\mu}$, in order to generate an appropriate repulsive effect to obtain a stable nucleus.  The coupling constant $g_{\omega}$ is therefore considered to be a free parameter.  The omega meson acquires its mass by chiral symmetry breaking as the case of the nucleon field \cite{boguta1,ogawa1}. The field strength of the isoscalar vector meson is given by
\begin{equation}
\Omega_{\mu\nu}=\partial_{\mu}\omega_{\nu}-\partial_{\nu}\omega_{\mu}.
\end{equation}
Up to this stage this Lagrangian has a mass term $\mu^2$ of scalar bosons. We take the limit $\mu^2\rightarrow 0$ in the Coleman-Weinberg renormalization procedure. In this sense we name it the massless chiral sigma model (MCSM). We describe $\phi$ for the scalar field {\it before} the chiral symmetry breaking and $\sigma$ {\it after} the chiral symmetry breaking hereafter.

\subsection{\label{subsec:level1}Boson loop with chiral symmetry} 

\begin{figure}
\includegraphics[scale=0.6]{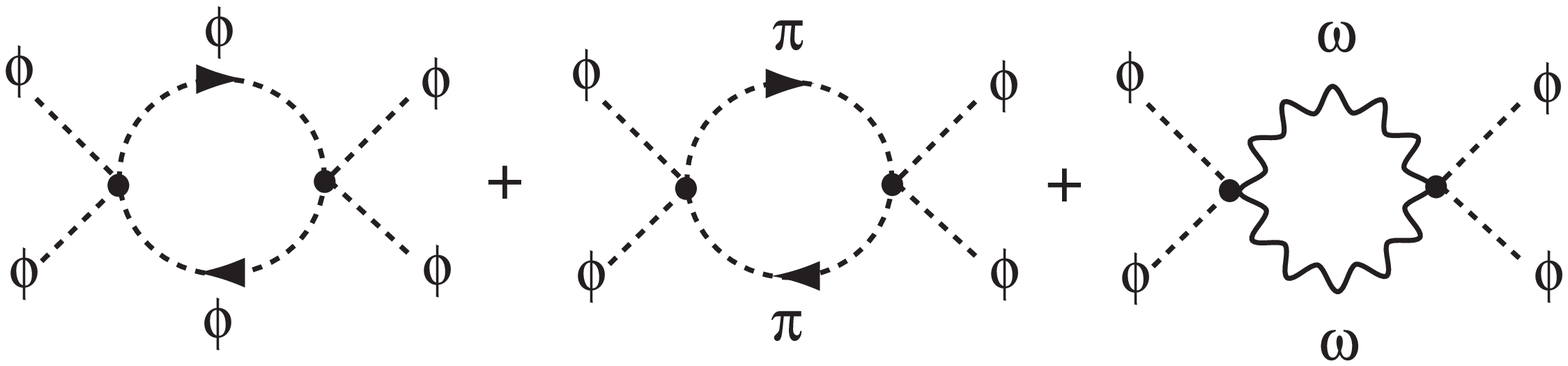}
\caption{\label{fig:fig2}For the four external lines of sigma meson,
for example, three-type diagrams exist.}
\end{figure}

We expect that the boson loop effective potential with chiral symmetry is almost the same as the one in the previous section. It is necessary to consider new diagrams with different bosons as shown in Fig.~\ref{fig:fig2}.
In order to respect the chiral symmetry we must deal with 
sigma and pi meson equally. The effective action becomes
\begin{eqnarray}
\Gamma_B^{\chi}&=&\int d^4x\left[\frac{i}{2}\ln\det\left(-\frac{\delta^2\mathcal{L}}{\delta b \delta b}\right)-VEV-\delta\mathcal{L}_{CTC}^B\right]\nonumber\\
{}&=&\int d^4x \left[-V_B^R(\phi,\bm\pi)+\frac{1}{2}Z_{\sigma\pi}^B(\partial_{\mu}\phi\partial^{\mu}\phi+\partial_{\mu}{\bm \pi}\cdot\partial^{\mu}{\bm \pi})+\cdots\right]\label{eq:baction},
\end{eqnarray}
where $\mathcal{L}$ includes the tree contribution and $\delta\mathcal{L}^B_{CTC}$ is the counterterms for the boson loop.
$\frac{\delta^2\mathcal{L}}{\delta b\delta b}$ means the second functional derivatives 
of $\mathcal{L}$ with respect to bosons (sigma, pi, and omega mesons).
Here we can neglect the external lines of vector meson owing to the current conservation.
For simplicity we introduce the new variable,
\begin{eqnarray}
\rho^2=\phi^2+{\bm \pi}^2.
\end{eqnarray}
We change the renormalization conditions for Eq.~(\ref{eq:baction})
using its variable as 
\begin{eqnarray}
\left.\frac{\partial^2 V_B^R}{\partial\rho^2}\right|_{\rho^2=0}&=&0,\\
{}\left.\frac{\partial^4 V_B^R}{\partial\rho^4}\right|_{\rho^2=m^2}&=&0,\\
{}\left.Z_{\sigma\pi}^B\right|_{\rho^2=m^2}&=&0.
\end{eqnarray}

Finally the renormalized potential of boson with the chiral symmetry becomes 
\begin{eqnarray}
V_B^R&=&\frac{\rho^4}{256\pi^2}\left[(6\lambda)^2\left\{1+3\left(\frac{1}{3}\right)^2\right\}+12\widetilde{g_{\omega}}^4\right]\left[\ln\left(\frac{\rho^2}{m^2}\right)-\frac{25}{6}\right]\nonumber\\
{}&=&\frac{3\rho^4}{64\pi^2}\left(4\lambda^2+\widetilde{g_{\omega}}^4\right)\left[\ln\left(\frac{\rho^2}{m^2}\right)-\frac{25}{6}\right]\nonumber\\
{}&=&\frac{3\left(4\lambda^2+\widetilde{g_{\omega}}^4\right)}{64\pi^2}(\phi^2+{\bm \pi}^2)^2\left[\ln\left(\frac{\phi^2+{\bm \pi}^2}{m^2}\right)-\frac{25}{6}\right]\label{eq:bloop}.
\end{eqnarray}
The calculation is entirely analogous to Eq.~(\ref{eq:rad}).
We only need to note that the $\phi^2{\bm\pi}^2$ coupling is
$1/3$ of the $\phi^4$ coupling and there are three kinds of 
pion and that the extra factor 12
in the omega loop comes from the trace of Lorentz-gauge propagator
and coupling constant. In the same way we calculate the coefficients of
derivative terms by including all contributions from boson and find before the renormalization,
\begin{eqnarray}
Z_{\sigma\pi}^B=\frac{2\lambda+\widetilde{g_{\omega}}^2}{16\pi^2}.
\end{eqnarray}
This constant correction is removed by taking a counterterm to cancel out.  Hence, we have $Z_{\sigma\pi}^B=0$ eventually.

\subsection{\label{subsec:level2}Nucleon loop with chiral symmetry} 

It is important to take into account the nucleon loop
before the chiral symmetry breaking
since the chiral symmetry is fulfilled only in the massless phase.
However we encounter the same difficulty
like the logarithmic singularity of boson loop in the massless limit.
At first we calculate the nucleon loop with finite mass
in the same way as the boson loop.
Then we take the limit of massless phase ($M\rightarrow 0$) and
replace $\sigma$ with $\phi$ at the same time in the renormalization procedure.
The one-loop effective action of nucleon with chiral symmetry is given by
\begin{eqnarray}
\Gamma_F^{\chi}&=&\int d^4x\left[-i\tr\ln\left\{\slash{k}-M-g_{\sigma}(\sigma+i\gamma_5\bm{\tau}\cdot\bm{\pi})-g_{\omega}\gamma_{\mu}\omega^{\mu}\right\}-VEV-\delta\mathcal{L}^F_{CTC}\right]\nonumber\\
{}&=&\int d^4x\left[-V_F^R(\phi,\bm\pi)+\frac{1}{2}Z_{\sigma\pi}^F(\partial_{\mu}\phi\partial^{\mu}\phi+\partial_{\mu}{\bm \pi}\cdot\partial^{\mu}{\bm \pi})+\frac{1}{4}Z_{\omega}^F\Omega_{\mu\nu}\Omega^{\mu\nu}+\cdots\right]\label{eq:naction},
\end{eqnarray}
where $\delta\mathcal{L}^F_{CTC}$ is the counterterms for nucleon loop.
\begin{figure}
\includegraphics[scale=0.6]{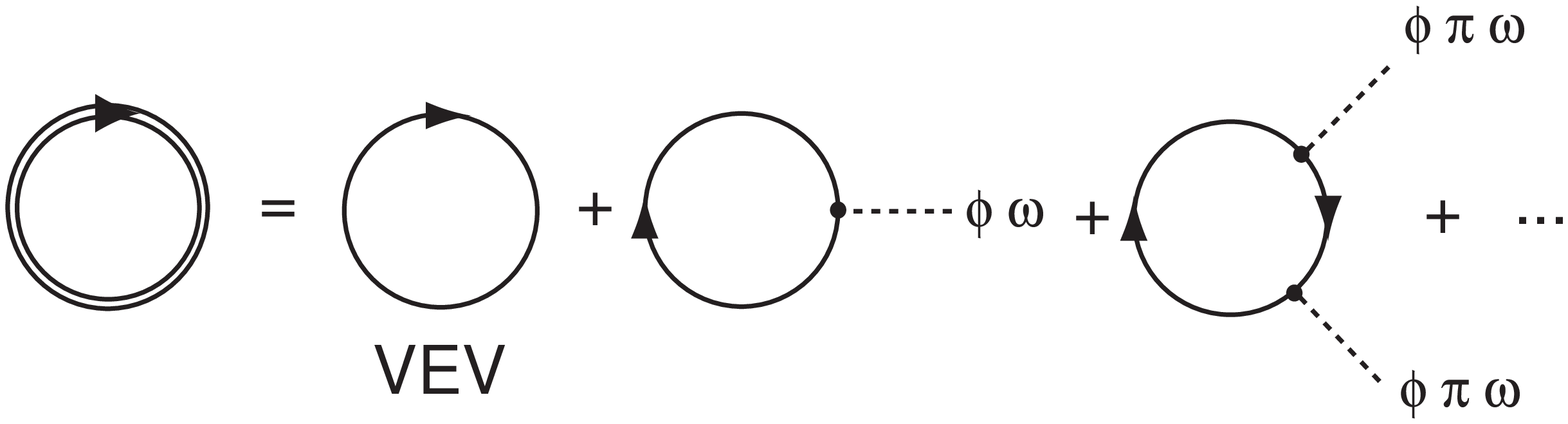}
\caption{\label{fig:fig3}Nucleon loop potential in the relativistic mean field (RMF) approximation. Double line is nucleon propagator with RMF and includes
all-order scalar, pseudoscalar, and vector potentials from Dyson equation.
Note that the contributions of nucleon loops coupled with odd-order pion
couplings vanish due to the property of $\gamma_5$. First term in right hand side means
vacuum expectation value (VEV) corresponding to 
the loop contribution from the valence nucleon.}
\end{figure}
In the calculation it is important to consider the
one-nucleon loop contribution in the massless phase.
In the phase of the chiral symmetry breaking the diagrams
from sigma meson are different from ones from pi meson
due to the property of pseudoscalar coupling in Fig.~\ref{fig:fig3}.
However we can deal with both sigma and pi mesons symmetrically
before the chiral symmetry breaking.
We take the renormalization conditions for Eq.~(\ref{eq:naction})
using new variable
\begin{eqnarray}
\left.\frac{\partial^2 V_F^R}{\partial\rho^2}\right|_{\rho^2=0}&=&0,\\
{}\left.\frac{\partial^4 V_F^R}{\partial\rho^4}\right|_{\rho^2=m^2}&=&0,\\
{}\left.Z_{\sigma\pi}^F\right|_{\rho^2=m^2}&=&0,\\
{}\left.Z_{\omega}^F\right|_{\rho^2=m^2}&=&0.
\end{eqnarray}
Using these conditions we obtain the renormalized potential of nucleon loop as
\begin{eqnarray}
V_F^R&=&-\frac{g_{\sigma}^4\rho^4}{8\pi^2}\left[\ln\left(\frac{\rho^2}{m^2}\right)-\frac{25}{6}\right]\nonumber\\
{}&=&-\frac{g_{\sigma}^4}{8\pi^2}(\phi^2+{\bm \pi}^2)^2\left[\ln\left(\frac{\phi^2+{\bm \pi}^2}{m^2}\right)-\frac{25}{6}\right]\label{eq:nloop},\\
{}Z_{\sigma\pi}^F&=&-\frac{g_{\sigma}^2}{4\pi^2}\ln\left(\frac{\phi^2+\bm\pi^2}{m^2}\right),\\
{}Z_{\omega}^F&=&\frac{g_{\omega}^2}{6\pi^2}\ln\left(\frac{\phi^2+\bm\pi^2}{m^2}\right).
\end{eqnarray}
In the same way as the boson loop, we also introduce the same renormalization scale $m$ 
in order to avoid a logarithmic singularity. 
As one can see in Eqs. (\ref{eq:bloop}) and (\ref{eq:nloop}), the difference between boson and fermion loops is sign and coupling constant, but both of them have the same function forms.
This is true before chiral symmetry breaking and by using 
this renormalization procedure we can deal with 
the loop contributions from the boson and nucleon symmetrically.
Through these good features we define the absolute ratio of $V_B^R$ to $V_F^R$ in order to estimate the loop corrections from boson and nucleon,
\begin{eqnarray}
\gamma=\left|\frac{V_B^R}{V_F^R}\right|=\frac{3\left(4\lambda^2+\widetilde{g_{\omega}}^4\right)}{8g_{\sigma}^4}\label{eq:ratio}.
\end{eqnarray}
It is possible to obtain the total renormalized potential using this ratio as 
\begin{eqnarray}
V_{all}^R=V_B^R+V_F^R=\frac{\gamma -1}{8\pi^2}g_{\sigma}^4(\phi^2+{\bm \pi}^2)^2\left[\ln\left(\frac{\phi^2+{\bm \pi}^2}{m^2}\right)-\frac{25}{6}\right]\label{eq:totalloop}.
\end{eqnarray}

\subsection{\label{subsec:level3}Total Lagrangian and spontaneous chiral symmetry breaking} 
In the above two subsections, we have obtained the renormalized one-loop potential of boson and nucleon with the chiral symmetry. 
The massless chiral sigma model with Eq.~(\ref{eq:totalloop}) becomes
\begin{eqnarray}
\mathcal{L}^{MCSM}&=&\bar{\psi}\left[i\gamma_{\mu}\partial^{\mu}-g_{\sigma}(\phi+i\gamma_5{\bm \tau}\cdot{\bm \pi})-g_{\omega}\gamma_{\mu}\omega^{\mu}\right]\psi\nonumber\\
{}&+&\frac{1}{2}\left(\partial_{\mu}\phi\partial^{\mu}\phi+\partial_{\mu}{\bm \pi}\cdot\partial^{\mu}{\bm \pi}\right)-\frac{\lambda}{4}(\phi^2+{\bm \pi}^2)^2\nonumber\\
{}&-&\frac{1}{4}\Omega_{\mu\nu}\Omega^{\mu\nu}+\frac{1}{2}\widetilde{g_{\omega}}^2(\phi^2+{\bm \pi}^2)\omega_{\mu}\omega^{\mu}\nonumber\\
{}&-&\frac{\gamma-1}{8\pi^2}g_{\sigma}^4(\phi^2+{\bm \pi}^2)^2\left[\ln\left(\frac{\phi^2+{\bm \pi}^2}{m^2}\right)-\frac{25}{6}\right]\nonumber\\
{}&+&\frac{1}{2}Z_{\sigma\pi}\left(\partial_{\mu}\phi\partial^{\mu}\phi+\partial_{\mu}{\bm \pi}\cdot\partial^{\mu}{\bm \pi}\right)+\frac{1}{4}Z_{\omega}\Omega_{\mu\nu}\Omega^{\mu\nu}\nonumber\\
{}&+&\epsilon\phi.
\end{eqnarray}
Here, we have added the explicit chiral symmetry breaking term, $\epsilon\phi$, which produces the finite pion mass after chiral symmetry breaking.  The functional coefficients of the derivative terms are given by
\begin{eqnarray}
Z_{\sigma\pi}&=&Z_{\sigma\pi}^F+Z_{\sigma\pi}^B=-\frac{g_{\sigma}^2}{4\pi^2}\ln\left(\frac{\phi^2+{\bm \pi}^2}{m^2}\right),\label{eq:sfunc}\\
{}Z_{\omega}&=&Z_{\omega}^F=\frac{g_{\omega}^2}{6\pi^2}\ln\left(\frac{\phi^2+{\bm \pi}^2}{m^2}\right).
\end{eqnarray}
Here, we have defined a non-trivial local minimum away from the origin
using $<0|\phi|0>=f_{\pi}$ and $<0|\pi|0>=0$ at the zero density (vacuum), together with $<0|\omega_{\mu}|0>=0$ as
\begin{eqnarray}
\left.\frac{\partial U_{all}}{\partial \phi}\right|_{\phi=f_{\pi},{\bm \pi}=0,\omega_{\mu}=0}=0.\label{eq:local}
\end{eqnarray}
where $U_{all}$ means all of the tree and loop contributions.
Eq.~(\ref{eq:local}) for the local minimum determines the coupling
constant $\lambda$ dependent on the renormalization scale $m$,
\begin{eqnarray}
\frac{3}{2\pi^2}\left[\ln\left(\frac{f_{\pi}}{m}\right)-\frac{11}{6}\right]\lambda^2+\lambda-\frac{g_{\sigma}^4-\frac{3}{8}\widetilde{g_{\omega}}^4}{\pi^2}\left[\ln\left(\frac{f_{\pi}}{m}\right)-\frac{11}{6}\right]-\frac{\epsilon}{f_{\pi}^3}=0.\label{eq:min}
\end{eqnarray}
Eq.~(\ref{eq:min}) has two solutions as a function of $m$ and we choose the positive coupling constant $\lambda$ as the natural choice.
\begin{figure}
\includegraphics[scale=0.6]{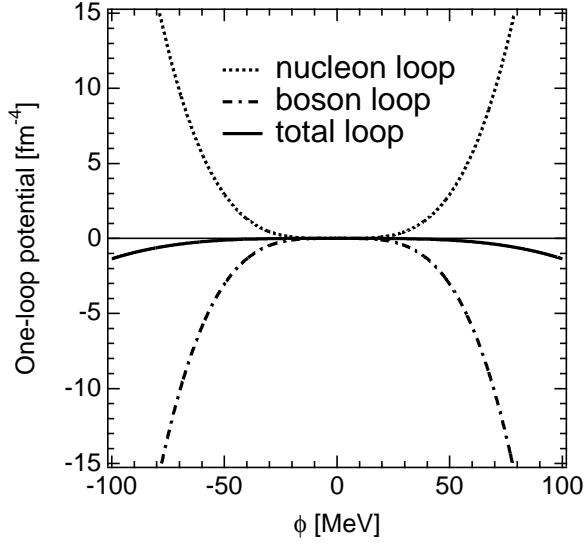}
\caption{\label{fig:fig4}One-loop potentials from nucleon and bosons
(Eqs.~(\ref{eq:bloop}) and (\ref{eq:nloop}))
as a function of the field $\phi$ of sigma meson
before chiral symmetry breaking with Eq.~(\ref{eq:min}).  We take $m=f_{\pi}$ for this presentation with the values of the coupling constants in Table I.
The total effective potential (Eq.~(\ref{eq:totalloop})) is negative as denoted by the solid curve.
This property provides spontaneous chiral symmetry breaking.}
\end{figure}
As shown in Fig.~\ref{fig:fig4} both the boson and the nucleon loops are too large as compared with the tree contributions. However, the total loop potential is a reasonable and negative one due to cancellation between the large positive potential
from the nucleon loop and the large negative one from the boson loop.
As a result, the total renormalized loop potential plays an important role as the negative mass term of the linear sigma model and the source of the Higgs mechanism through Eq.~(\ref{eq:local}).
 
In the linear representation of chiral symmetry we obtain the 
Lagrangian as 
\begin{eqnarray}
\mathcal{L}&=&\bar{\psi}\left[i\gamma_{\mu}\partial^{\mu}-M-g_{\sigma}(\sigma+i\gamma_5{\bm \tau}\cdot{\bm \pi})-g_{\omega}\gamma_{\mu}\omega^{\mu}\right]\psi\nonumber\\
{}&+&\frac{1}{2}\left(1+Z_{\sigma\pi}'\right)\left(\partial_{\mu}\sigma\partial^{\mu}\sigma+\partial_{\mu}{\bm \pi}\cdot\partial^{\mu}{\bm \pi}\right)\nonumber\\
{}&-&\frac{3}{2}\lambda f_{\pi}^2\sigma^2-\lambda f_{\pi}\sigma^3-\frac{1}{4}\lambda\sigma^4-\frac{1}{2}\lambda f_{\pi}^2{\bm \pi}^2-\lambda f_{\pi}\sigma{\bm \pi}^2-\frac{1}{2}\lambda\sigma^2{\bm \pi}^2-\frac{1}{4}\lambda{\bm \pi}^4\nonumber\\
{}&-&\frac{1}{4}\left(1-Z_{\omega}'\right)\Omega_{\mu\nu}\Omega^{\mu\nu}+\frac{1}{2}m_{\omega}^2\omega_{\mu}\omega^{\mu}+\frac{1}{2}\widetilde{g_{\omega}}^2(2f_{\pi}\sigma+\sigma^2+{\bm \pi}^2)\omega_{\mu}\omega^{\mu}\nonumber\\
{}&-&\frac{\gamma-1}{8\pi^2}g_{\sigma}^4\left[\left(f_{\pi}^{*2}+\bm\pi^2\right)^2\ln\left(\frac{f_{\pi}^{*2}+\bm\pi^2}{m^2}\right)-2f_{\pi}^4\ln\left(\frac{f_{\pi}}{m}\right)\right.\nonumber\\
{}&-&\left.\frac{25}{6}\left\{\left(f_{\pi}^{*2}+\bm\pi^2\right)^2-f_{\pi}^4\right\}-4\sigma f_{\pi}^3\left\{2\ln\left(\frac{f_{\pi}}{m}\right)-\frac{11}{3}\right\}\right]\label{eq:linear},
\end{eqnarray}
where 
\begin{eqnarray}
M&=&g_{\sigma}f_{\pi}, \label{eq:nmass}\\
{}f_{\pi}^*&=&f_{\pi}+\sigma,\\
{}M^*&=&M+g_{\sigma}\sigma=g_{\sigma}f_{\pi}^*, \\
{}m_{\omega}&=&\widetilde{g_{\omega}}f_{\pi}, \label{eq:omass}\\
{}Z_{\sigma\pi}'&=&-\frac{g_{\sigma}^2}{4\pi^2}\ln\left(\frac{f_{\pi}^{*2}+\bm \pi^2}{m^2}\right),\\
{}Z_{\omega}'&=&\frac{g_{\omega}^2}{6\pi^2}\ln\left(\frac{f_{\pi}^{*2}+\bm\pi^2}{m^2}\right).
\end{eqnarray}
The spontaneous chiral symmetry breaking makes nucleon,
sigma meson, and omega meson massive.
Only the pion mass is generated from the explicitly chiral symmetry
breaking term. The masses of scalar and pseudoscalar mesons are given 
from the effective potential by
\begin{eqnarray}
m_{\sigma}^{2}&=&\left.\frac{\partial^2 U_{all}}{\partial \sigma^2}\right|_{\sigma=0,{\bm \pi}=0}=3\lambda f_{\pi}^2+(\gamma -1)\frac{3M^2g_{\sigma}^2}{2\pi^2}\left[2\ln\left(\frac{f_{\pi}}{m}\right)-3\right],\label{eq:smass}\\
m_{\pi}^{2}&=&\left.\frac{\partial^2 U_{all}}{\partial {\bm \pi}^2}\right|_{\sigma=0,{\bm \pi}=0}=\frac{\epsilon}{f_{\pi}}\label{eq:pmass}.
\end{eqnarray}
We take the masses and the pion decay constant as $M=$ 939 [MeV], 
$m_{\omega}=$ 783 [MeV], $m_{\pi}=$ 139 [MeV], and $f_{\pi}=$ 93 [MeV].
Then, the other parameters can be determined automatically using Eqs.~(\ref{eq:min}), (\ref{eq:nmass}), (\ref{eq:omass}), (\ref{eq:smass}), and (\ref{eq:pmass}) in Table~\ref{tab:para}.  
\begin{table}
\begin{center}
\caption{{\bf {\small Parameter sets
through the relationships using $m=f_{\pi}$. \label{tab:para}}}}
\begin{tabular}{cccc|cccccc}
\hline
\hline
$M$[MeV] & $m_{\omega}$[MeV] & $m_{\pi}$[MeV] & $f_{\pi}$[MeV] & $g_{\sigma}$ &
$\widetilde{g_{\omega}}$ & $\lambda$ & $\gamma$ & $\epsilon$[MeV$^3$] & $m_{\sigma}$[MeV] \\
\hline
939  & 783  & 139  &  93   & 10.09 & 8.419 &
77.07 & 1.038 & 1.79$\times 10^6$  & 641  \\
\hline
\hline
\end{tabular}
\end{center}
\end{table}

\begin{figure}
\includegraphics[scale=0.6]{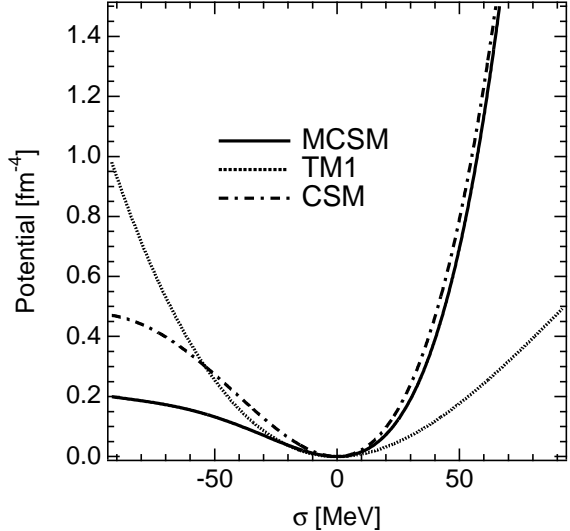}
\caption{\label{fig:fig5}The effective potential as a
function of the new field $\sigma$ of sigma meson 
around the new local minimum using the massless chiral sigma model 
(MCSM) by the solid curve at $m=f_{\pi}$. 
As a comparison, the tree level potential of the chiral sigma model
\cite{ogawa1} and that of the TM1 parameter set 
\cite{sugahara} are represented by
the dash-dotted and dotted curves, respectively.}
\end{figure}
Using all of the parameters we plot the effective potential around
the new local minimum as a function of the scalar field $\sigma$ 
in Fig.~\ref{fig:fig5}. The renormalized effective potential consistent with chiral symmetry becomes stable around the new origin 
satisfying Eq.~(\ref{eq:min}) by using the Coleman-Weinberg renormalization procedure for the first time.
It is amazing that the results in Fig.~\ref{fig:fig4} and \ref{fig:fig5} come out without free parameters. These results come out naturally from the chiral symmetry in the Coleman-Weinberg scheme.

\section{\label{sec:level4}Naturalness in the massless chiral sigma model}

We estimate here the validity of the vacuum contribution using the
naive dimensional analysis (NDA) \cite{georgi}.
Although the Walecka model and the chiral sigma model
in the linear representation are renormalizable,
it is known that the effects of not only chirally symmetric renormalization
using ordinary procedure 
but also one-loop potential from nucleon in the Walecka model
are not natural by the NDA \cite{furn,friar}.

In this section we follow the definition and 
the convention for the NDA in Ref.~\cite{furn}.
For example a term in the scalar effective potential takes the form
using the appropriate dimensional scales
\begin{eqnarray}
U_n(\sigma)=\kappa_n\frac{1}{n!}f_{\pi}^2\Lambda^2\left(\frac{\sigma}{f_{\pi}}\right)^n\label{eq:scale}.
\end{eqnarray}
The dimensionless coefficients $\kappa_n$ should be of order unity
if naturalness holds.
In this paper we use $\Lambda = M= 939$ [MeV].

At first we evaluate the naturalness in the Walecka model as mentioned above.
The contribution of nucleon loop is written as \cite{serot,chin}
\begin{eqnarray}
{}& &\Delta\mathcal{E}_{VF}^{Walecka}\nonumber\\
{}&=&-\frac{1}{4\pi^2}\left[M^{*4}\ln\left(\frac{M^*}{M}\right)-M^3\Phi-\frac{7}{2}M^2\Phi^2-\frac{13}{3}M\Phi^3-\frac{25}{12}\Phi^4\right]\nonumber\\
{}&=&-\frac{M^4}{4\pi^2}\left[\frac{1}{5}\left(\frac{\Phi}{M}\right)^5-\frac{1}{30}\left(\frac{\Phi}{M}\right)^6+\cdots+\frac{4!(n-5)!(-1)^{n-1}}{n!}\left(\frac{\Phi}{M}\right)^n\right]\label{eq:vacwalecka}.
\end{eqnarray}
where $\Phi=g_{\sigma}\sigma$.
The leading term of Eq.~(\ref{eq:vacwalecka}) should be scaled based on the scaling
rules (Eq.~(\ref{eq:scale})) as 
\begin{eqnarray}
-\frac{M^4}{4\pi^2}\frac{g_{\sigma}^5\sigma^5}{5M^5}=\kappa_5\left(\frac{M^2}{5!f_{\pi}^3}\sigma^5\right)\rightarrow \kappa_5=-61.97
\end{eqnarray}
Thus the one-nucleon loop contribution to the energy has a large naturalness value. However, total effective potential with tree contributions and one-loop corrections are still stable in the Walecka model.

Next we consider naturalness of the massless chiral sigma model in the NDA.
The vacuum fluctuation (Eq.~(\ref{eq:linear})) of scalar meson becomes
\begin{eqnarray}
\Delta\mathcal{E}_{VF}(\sigma)_{\pi=0}&=&\frac{\gamma -1}{8\pi^2}g_{\sigma}^4\left[2f_{\pi}^{*4}\ln\left(\frac{f_{\pi}^*}{m}\right)-2f_{\pi}^4\ln\left(\frac{f_{\pi}}{m}\right)-\frac{25}{6}\left(f_{\pi}^{*4}-f_{\pi}^4\right)\right.\nonumber\\
{}& &\left.-4\sigma f_{\pi}^3\left\{2\ln\left(\frac{f_{\pi}}{m}\right)-\frac{11}{3}\right\}\right]\nonumber\\
{}&=&\frac{\gamma -1}{4\pi^2}M^4\left[-9\left(\frac{\Phi}{M}\right)^2-4\left(\frac{\Phi}{M}\right)^3+\frac{1}{5}\left(\frac{\Phi}{M}\right)^5-\frac{1}{30}\left(\frac{\Phi}{M}\right)^6\right.\nonumber\\
{}& &\left.+\cdots +\frac{4!(n-5)!(-1)^{n-1}}{n!}\left(\frac{\Phi}{M}\right)^n\right],\label{eq:vacmcsm}
\end{eqnarray}
where we choose $m=f_{\pi}$.
Especially a leading term of Eq.~(\ref{eq:vacmcsm}) 
and another leading term which comes from
radiative corrections are evaluated in the NDA
\begin{eqnarray}
-\frac{\gamma -1}{4\pi^2}\frac{9M^4\Phi^2}{M^2}&=&\kappa_2\left(\frac{M^2}{2!}\sigma^2\right)\rightarrow \kappa_2=-\frac{9g_{\sigma}^2}{2\pi^2}(\gamma -1)=-1.76,\\
{}\frac{\gamma -1}{4\pi^2}\frac{M^4\Phi^5}{5M^5}&=&\kappa_5\left(\frac{M^2}{5!f_{\pi}^3}\sigma^5\right)\rightarrow \kappa_5=\frac{3!g_{\sigma}^2}{\pi^2}(\gamma -1)=2.35.
\end{eqnarray}
When we consider the effect of only nucleon loop in any model and any renormalization scheme, it has too large non-linear potential and unnatural coefficient. By introducing both nucleon and boson loops before the symmetry breaking in the Coleman-Weinberg scheme, we find that two contributions from vacuum fluctuation are almost cancelled. In the RMF consistent with chiral symmetry, the naturalness is fulfilled even with the introduction of one-loop corrections of nucleon and bosons for the first time.

\section{\label{sec:level5}The physical meaning of the renormalization scale}
\begin{figure}
\includegraphics[scale=0.6]{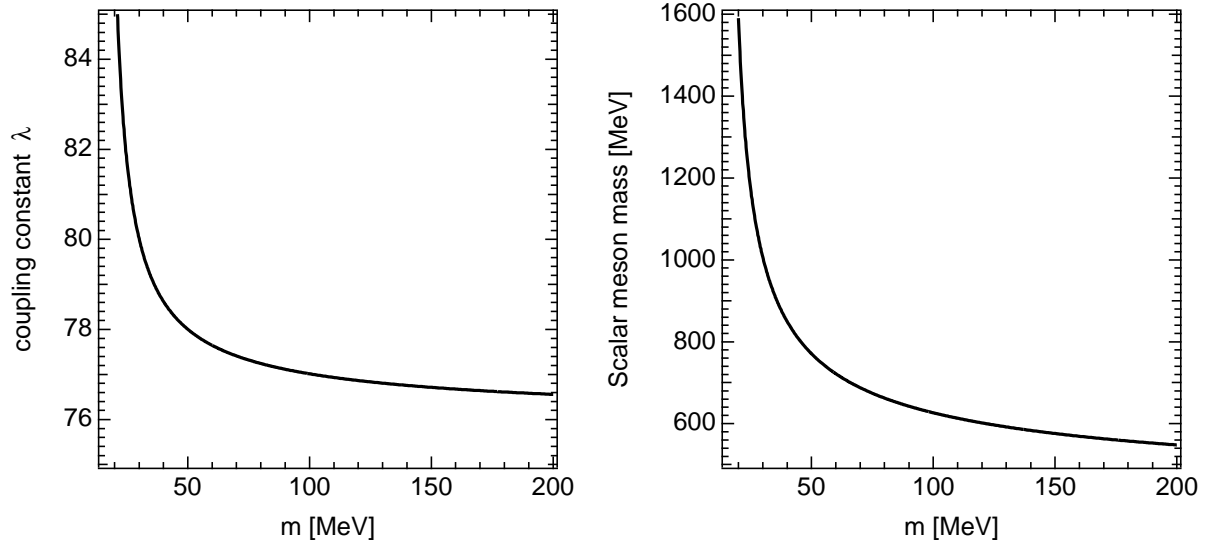}
\caption{The dependence of the coupling constant 
$\lambda$ and the mass of the sigma meson on $m$.}
\label{fig:fig6}
\end{figure}
\begin{figure}
\includegraphics[scale=0.6]{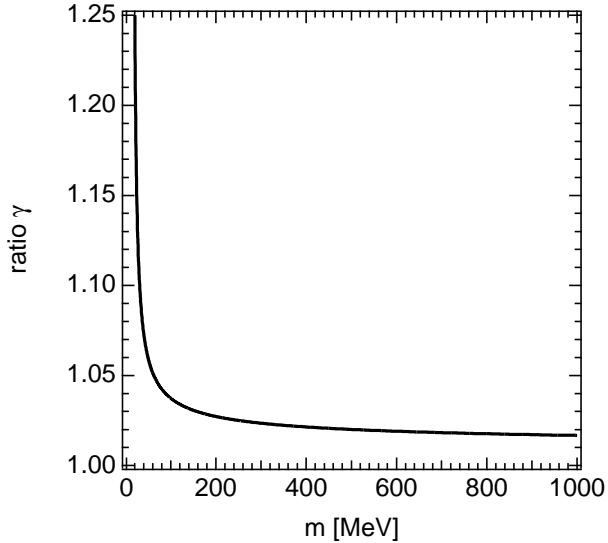}
\caption{The dependence of the ratio $\gamma$ on $m$.
This figure shows that the contribution from the boson loop is larger than the one from the nucleon loop. It corresponds to the result of Fig.~\ref{fig:fig4}. Taking the limit $m\rightarrow\infty$ with Eq.~(\ref{eq:min}), this ratio goes to 1. It means that the one-loop corrections from boson and nucleon perfectly cancel each other in this limit.}
\label{fig:fig7}
\end{figure}

We consider here the physical meaning of the renormalization scale $m$ in the massless chiral sigma model through the ratio $\gamma$.
At first we consider the coupling constant $\lambda$ (Eq.~(\ref{eq:min}))
and the mass of the sigma meson $m_{\sigma}$ (Eq.~(\ref{eq:smass})) for their dependence on the renormalization scale $m$ in Fig.~\ref{fig:fig6}. In the small $m$ region,
the dependence of two values on $m$ is large, 
while it saturates in the large $m$ region.
We find an interesting property of dependence on renormalization 
scale in the limit $m\rightarrow \infty$
\begin{eqnarray}
\lambda&\rightarrow&75.315,\\
m_{\sigma}&\rightarrow&0.\label{eq:mass0}
\end{eqnarray}
Especially Eq.~(\ref{eq:mass0}) shows the restoration of the 
chiral symmetry in the limit of $m\rightarrow \infty$. Since one-loop corrections give rise to spontaneous chiral symmetry breaking
in our model, it means that this generator vanishes in the limit $m\rightarrow\infty$.
In order to check this fact we estimate the ratio (Eq.~(\ref{eq:ratio})) with
Eq.~(\ref{eq:min}),
\begin{eqnarray}
\gamma\rightarrow 1\ \ (m\rightarrow\infty).\label{eq:gamma0}
\end{eqnarray}
As shown in Fig.~\ref{fig:fig7}, all of the radiative corrections
are perfectly cancelled in this limit, and 
the mechanism of spontaneous symmetry breaking does not occur.
Thus both nucleon and boson becomes massless particle in this phase. 
Once a renormalization scale has a finite value, the symmetry between fermion and boson is broken. 
At the same time, spontaneous symmetry breaking occurs and 
the masses of both nucleon and bosons are generated dynamically.
This means that the MCSM becomes trivial for $m=\infty$, while the system becomes non-trivial at finite $m$.
It is now legitimate to set $m=f_\pi$, since the renormalization condition is naturally taken at the point where the nucleon and all the bosons should have the known physical values in the non-trivial vacuum.

\section{\label{sec:level6}Conclusion}

We have studied the vacuum polarization of the Dirac sea with chiral symmetry using the Coleman-Weinberg renormalization procedure. In the Coleman-Weinberg scheme, we start with the massless fermion and boson Lagrangian with the chiral symmetry, where the renormalization scale is chosen at finite $m$.  We work out the renormalization procedure and obtain a divergence-free chiral Lagrangian, which includes the vacuum polarization effect.  The mean field approximation provides the lowest energy state as the ground state, which corresponds to the sigma field $\phi=f_\pi$.  We take the renormalization scale at $m=f_\pi$, because this is the vacuum state, where the nucleon mass and meson masses are known as the experimental values.  Hence, all the parameters in the Lagrangian are fixed by the experiment.

It is interesting to analyze the Coleman-Weinberg renormalization scheme applied to the fermion-boson system with chiral symmetry.  The contributions of the loop diagrams of both the fermion and boson come out to have the same functional forms except for the coupling constants.  The renormalized Lagrangian at some renormalization scale $m$ provides non-trivial solution with the lowest mean field energy at $\phi=f_\pi$ for the case of the finite renormalization scale.  However, as $m \rightarrow \infty$ the fermion loop and the boson loop cancel completely each other and the model becomes trivial.  In this limit the quantum corrections completely disappears.

We have now a theoretical model to handle the chirally symmetric Lagrangian with the vacuum polarization effect having been worked out.  We are now armed to describe the nuclear system at finite nucleon density and also the hadronic system at finite temperature with the model Lagrangian with chiral symmetry.

\begin{acknowledgments}
We acknowledge fruitful discussions with Prof. A. Hosaka
on the renormalization and chiral symmetry.
This work is supported partially by the Sasakawa Scientific Research Grant from The Japan Science Society.
\end{acknowledgments}

\end{document}